\documentclass[aps,pr,reprint,twocolumn,groupedaddress]{revtex4}

\usepackage{latexsym}
\usepackage{amsthm}
\usepackage{amsmath}
\usepackage{graphicx}
\usepackage{mathtools}
\usepackage{slashed}
\usepackage{amssymb}
\usepackage{indentfirst}
\usepackage{subfigure}
\usepackage{color}
\usepackage{epsf}
\usepackage[colorlinks=False]{hyperref}

\newcommand*\diff{\mathop{}\!\mathrm{d}}
\newcommand*\Diff[1]{\mathop{}\!\mathrm{d^#1}}
\newcommand{\nn}{\nonumber}

\newcommand{\be}{\begin{eqnarray}}
\newcommand{\ee}{\end{eqnarray}}
\newcommand{\ma}{\mathrm}
\newcommand{\ml}{\mathcal}
\newcommand{\bs}{\boldsymbol}
\newcommand{\Tr}{\mathrm{Tr}}

\begin{document}

\title{Doubly charmed baryon production in heavy ion collisions}

\author{Xiaojun Yao}
\email{xiaojun.yao@duke.edu}
\author{Berndt M\"uller}
\email{mueller@phy.duke.edu}
\affiliation{Department of Physics, Duke University, Durham, NC 27708, USA}

\date{\today}

\begin{abstract}
We give an estimate of $\Xi_{cc}^{++}$ production rate and transverse momentum spectra in relativistic heavy ion collisions. We use Boltzmann transport equations to describe the dynamical evolution of charm quarks and diquarks inside quark-gluon plasma. In-medium formation and dissociation rates of charm diquarks are calculated from potential non-relativistic QCD for the diquark sector. We solve the transport equations by Monte Carlo simulations. For $2.76$ TeV Pb-Pb collisions with $0-10\%$ centrality, the number of $\Xi_{cc}^{++}$ produced in the transverse momentum range $0-5$ GeV and rapidity from $-1$ to $1$ is roughly $0.02$ per collision. We repeat the calculation with a melting temperature $250$ MeV above which no diquarks can be formed. The number of $\Xi_{cc}^{++}$ produced in the same kinematic region is about $0.0125$ per collision. We discuss how to study diquarks at finite temperature on a lattice and construct the anti-triplet free energy in a gauge invariant but path dependent way. We also comment on extensions of the calculation to other doubly heavy baryons and doubly heavy tetraquarks and the feasibility of experimental measurements.

\end{abstract}

\maketitle
Recently the LHCb Collaboration reported the observation of a doubly charmed baryon carrying two units of positive charge, $\Xi_{cc}^{++}$, with a mass $m(\Xi_{cc})\approx 3621$ MeV \cite{Aaij:2017ueg}. Though it is still unclear why the observed mass differs from the previous SELEX result \cite{Mattson:2002vu}, the existence of such doubly charmed baryons is now on a more solid ground. The particle is stable under strong interactions and only decays weakly. The structure of $\Xi_{cc}^{++}$ can be thought of as an up quark bound around a deeply bound state (diquark) of two charm quarks \cite{Karliner:2014gca}. Just as a pair of heavy quark and heavy anti-quark attract each other and can form a bound state in the color singlet channel, a pair of two heavy quarks also attract and can form a bound state, a heavy diquark, in the anti-triplet representation.

The peculiar properties of $\Xi_{cc}^{++}$ have stimulated new theoretical and experimental research. Here we consider the production of $\Xi_{cc}^{++}$ in high energy heavy ion collisions, where a hot nuclear environment, the quark-gluon plasma (QGP), is produced. Previous work was based on quark coalescence at hadronization and assumed that heavy quarks are thermally distributed  \cite{Becattini:2005hb, Zhao:2016ccp}. Here we pursue out a more dynamical approach considering the formation of bound heavy diquarks within the quark-gluon plasma and the incomplete equilibration of the heavy quark spectrum.

In hadron-hadron collisions, it is difficult to produce a pair of heavy quarks in the color anti-triplet at leading order in a fragmentation process. On the other hand, the coalescence process involving two independently produced charm quarks is sensitive to the relative momentum between the heavy quark pair. In proton-proton collisions, the relative momentum is uncontrolled and likely large, suppressing the coalescence. Heavy ion collisions have two advantages for $\Xi_{cc}^{++}$ production: First, the rapidity density of charm quarks produced in a single collision is higher. Second, the deconfined QGP medium lasts roughly $10$ fm/c, during which time the charm quarks can diffuse in the QGP via interactions with light quarks and gluons. This is confirmed by recent measurements from the STAR Collaboration, which shows that charm quarks participate in the collective flow of the QGP \cite{Adamczyk:2017xur}. 
As a result, the relative momentum of a charm quark pair can be on the order of the QGP temperature. The coalescence probability into a charm diquark bound state is thus enhanced if the temperature of the QGP is not too high. 

After its formation the charm diquark also diffuses in the QGP because it carries color charge. At the same time, the charm diquark may dissociate by absorbing a real or virtual gluon. So the whole process is a dynamical in-medium evolution involving charm diquark formation, diffusion and dissociation. This is similar to the in-medium evolution of heavy quarkonia, such as the $J/\psi$, except that the heavy diquarks carry color while the quarkonia are color neutral. At the transition from the deconfined QGP phase to the hadronic phase, the charm diquarks hadronize into doubly charmed baryons by absorbing an up or down quark from the medium.

We will describe the in-medium dynamical evolution of charm quarks and diquarks by a set of coupled Boltzmann equations analogous to the transport equations for in-medium heavy quarks and quarkonia \cite{Yao:2017fuc}. By connecting the transport equations with the initial production of charm quarks from the hard collision and the hydrodynamical background, we obtain an estimate of the yield and $p_T$-spectrum of $\Xi_{cc}^{++}$ in Pb-Pb collisions at $2.76$ TeV. Finally, we study the static screening effect of the QGP on the production process. 

The set of coupled Boltzmann transport equations for the charm quark and diquark distribution functions $f({\bs x}, {\bs p}, t)$ is given by
\be
\label{eq:LBE}
(\frac{\partial}{\partial t} + \dot{{\bs x}}\cdot \nabla_{\bs x})f_c({\bs x}, {\bs p}, t) &=& \ml{C}_c  - \ml{C}_c^+ + \ml{C}_c^-\\\nn
(\frac{\partial}{\partial t} + \dot{{\bs x}}\cdot \nabla_{\bs x})f_{cc}({\bs x}, {\bs p}, t) &=& \ml{C}_{cc} + \ml{C}_{cc}^+ - \ml{C}_{cc}^-\,,
\ee
where all the collision terms $\ml{C}$, $\ml{C}^{\pm}$ depend on ${\bs x}, {\bs p}, t$.
Here we will focus on the ground charm diquark state $cc$(1S) because excited states are loosely bound and cannot survive at high temperature. In the following, by charm diquark we mean the $cc$(1S) state.
The collision terms $\ml{C}_c$ and $\ml{C}_{cc}$ describe their scattering with thermal constituents of QGP. This process has been described as two-body scattering in the framework of the linearized Boltzmann equation~\cite{Gossiaux:2008jv,Gossiaux:2009mk,Uphoff:2014hza}. Here we use the elastic scattering rate calculated and implemented in Ref.~\cite{duke_lbt} to describe the in-medium diffusion. The diquark gain term $\ml{C}_{cc}^+$ is from the combination of a charm quark pair by gluon emission and the loss term $\ml{C}_{cc}^-$ is from dissociation by gluon absorption. The formation and dissociation of diquarks also change the charm quark distribution function, which are represented by $\ml{C}_{c}^{\pm}$.

We calculate the diquark formation and dissociation rates in QGP to the lowest order in potential non-relativistic QCD (pNRQCD) for the diquark sector \cite{Brambilla:2005yk,Fleming:2005pd}. The pNRQCD for the quarkonium sector has been used to study quarkonia dissociation rates inside QGP \cite{Brambilla:2008cx}. The effective field theory can be derived from QCD under the hierarchy of scales $M \gg Mv \gg Mv^2, T, m_D$ where $M=1.3$ GeV is the charm quark mass, $v\sim0.4$ is the relative velocity of $cc$ inside the diquark, $T$ is the QGP temperature, and $m_D$ is the Debye screening mass. If $T$ or $m_D$ scales as $Mv$, the Debye static screening of the color attraction is so strong that no diquark bound states can be formed inside QGP. So the above hierarchy of scales is relevant to the diquark formation. The pNRQCD is a systematic expansion in $v$ or $1/M$ (NR expansion) and $r$, the relative distance between the charm quark pair inside the diquark (multipole expansion). Its Lagrangian is given by: 
\begin{widetext}
\be
\ml{L}_{\ma{pNRQCD}} &=& \int \Diff{3}r \Tr \Big\{ \ma{T}^{\dagger} (iD_0 - H_T) \ma{T} 
+ \Sigma^{\dagger} (iD_0 - H_{\Sigma}) \Sigma 
+ \ma{T}^{\dagger} {\bs r} \cdot g {\bs E} \Sigma 
+ \Sigma^{\dagger} {\bs r} \cdot g {\bs E} \ma{T}\Big\} + \cdots\,,
\ee
\end{widetext}
where higher order interaction terms in $1/M$ and $r$ are omitted.
The Lagrangian of light quarks and gluons is just QCD with momenta $k \lesssim Mv$. The degrees of freedom are the anti-triplet $\ma{T}(\bs R, \bs r, t)$ and sextet $\Sigma(\bs R, \bs r, t)$ where $\bs R$ denotes the center-of-mass (c.m.) position and $\bs r$ the relative coordinate. They are defined as
\be
\ma{T} = t^l T^l \ \ \ \ \ \ \ \ \Sigma = \sigma^{\nu} \Sigma^{\nu} \,,
\ee
where $T^l$ and $\Sigma^{\nu}$ are the anti-triplet and sextet fields while $t^l $ and $\sigma^{\nu} $ are the generators of the corresponding representations. They are given by
\be
 t^l_{ij} &=& \frac{1}{\sqrt{2}}\epsilon_{ijl}\\
 \label{eqn:six1}
 \sigma^{1}_{11} &=& \sigma^{4}_{22}= \sigma^{6}_{33} = 1 \\
  \label{eqn:six2}
  \sigma^{2}_{12} &=&  \sigma^{2}_{21} =  \sigma^{3}_{13} =  \sigma^{3}_{31} =  \sigma^{5}_{23} =  \sigma^{5}_{32} = \frac{1}{\sqrt2}\,.
\ee
The equations of motion of the anti-triplet and sextet are Schr\"odinger equations with the Hamiltonians expanded in powers of $1/M$
\be
H_{T,\Sigma} &=& -\frac{{\bs D}_{\bs R} ^2 }{4M} - \frac{\nabla_{\bs r}^2}{M}  + V_{T,\Sigma}^{(0)} + \frac{V_{T,\Sigma}^{(1)}}{M} + \frac{V_{T,\Sigma}^{(2)}}{M^2} + \cdots\,,
\ee
where ${\bs D}_{\bs R}$ is the covariant derivative associated with the c.m. position.

By the virial theorem, $-\nabla_{\bs r}^2/M \sim V_{T,\Sigma}^{(0)}$. So the order of the relative kinetic term is accounted as $1/M^0$, not suppressed. The c.m. kinetic term is suppressed because momenta $k \sim Mv$ have been integrated out in the construction and then ${\bs D}_{\bs R}\ll Mv$. Higher-order terms of the potentials are also suppressed by $1/M$ which include relativistic corrections, spin-orbital and spin-spin interactions. We only work to order $1/M^0$ since the charm quark mass is large. At this order, the Hamiltonians only contain the relative kinetic term and $V_{T,\Sigma}^{(0)}$. Inside the deconfined QGP, the potential is flattened and can be approximated by Coulomb interactions
\be
V_{T}^{(0)} = -\frac{2}{3}\frac{\alpha_s}{r}\ \ \ \ \ \ \ \ V_{\Sigma}^{(0)} = \frac{1}{3}\frac{\alpha_s}{r}\,.
\ee
Since we keep track of the evolution of both the bound diquarks and unbound charm quarks in the Boltzmann equations, the potentials have no imaginary parts.

The interaction between the anti-triplet diquarks and the medium can be decomposed into two parts: a part that only changes the c.m. motion and leaves the bound state intact and the other part that only modifies the relative motion and can destroy the bound state. The decomposition is explicit in the pNRQCD Lagrangian by the multipole expansion. At the order we are working, the c.m. motion part is fully described by the gauged kinetic term of the anti-triplet field, in the same way as the interaction between heavy quarks and the medium. The changes of the c.m. motions of diquarks are treated as diffusion in the Boltzmann equation, in the same way as the heavy quark diffusion (see $\ml{C}_{c}$ and $\ml{C}_{cc}$ in expression (\ref{eq:LBE})). The change of the relative motion is described by terms of at least linear order in $r$. For example, the anti-triplet can interact with the sextet via a color dipole interaction where the chromoelectric field is given by
\be
{\bs E} = t^a_F {\bs E}^a\,,
\ee
and $t^a_F$ is the generator of the fundamental representation.

\begin{figure}
\centering
\vspace{0.1in}
\includegraphics[width=2.5in]{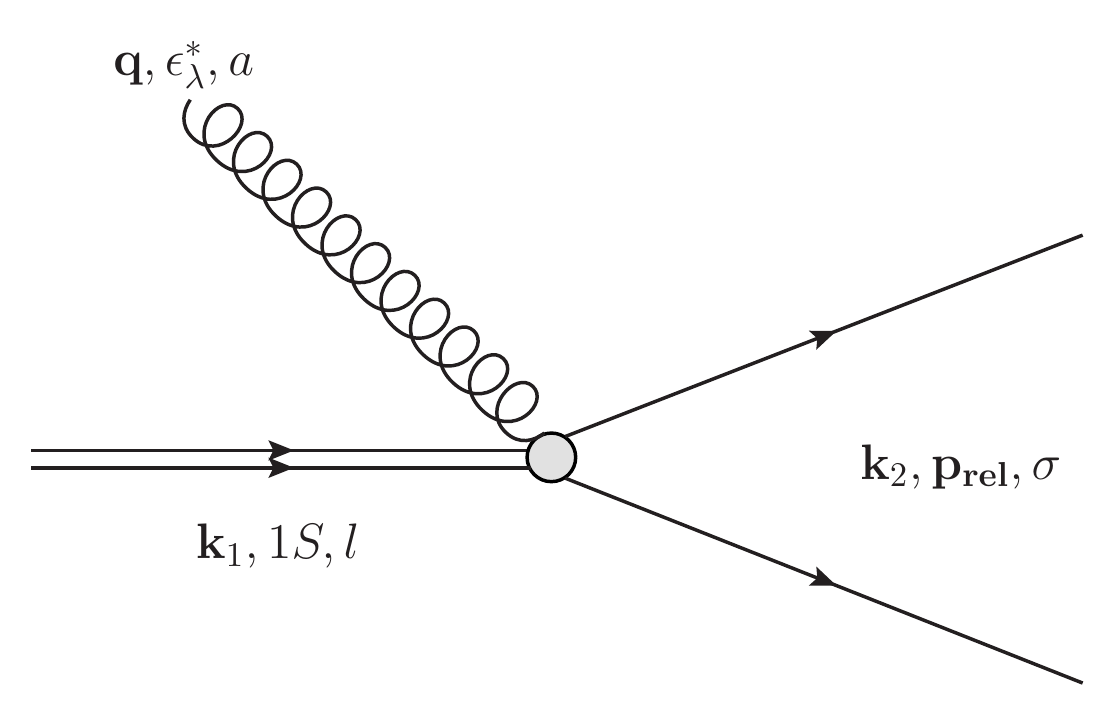}
\caption{Transition between a bound charm diquark in the anti-triplet and an unbound charm quark pair in the sextet by absorbing or emitting an on-shell gluon. Narrow double lines indicate the diquark while widely open double lines represent the unbound pair.}\label{fig:t_sigma}
\end{figure}

At leading order in $r$, the transition between unbound charm quark pairs and bound diquarks can only occur between an unbound sextet and a bound anti-triplet. 
The Feynman diagram of the transition via gluon absorption or emission is shown in Fig.~\ref{fig:t_sigma}. For simplicity, we only consider the interaction with on-shell gluons in the QGP. Transitions caused by virtual gluons (inelastic scattering with medium constitutes) are at next order in $\alpha_s$ and neglected here. The scattering amplitude in Coulomb gauge is given by 
\be
\ml{T}^{\nu l a}_\lambda &=& (2\pi)^4 \delta^3({\bs k}_1 + {\bs q} - {\bs k}_2) \delta(\Delta E) \ml{M}^{\nu l a}_\lambda\\
\ml{M}^{\nu l a}_\lambda &=& -igq \Tr( \sigma^{\nu}  t_F^a t^l ) (\epsilon_{\lambda}^*)_i \langle \psi_\ma{1S} | r_i | \Psi_{{\bs p}_\ma{rel}} \rangle \\
\Delta E &=& \frac{k^2_1}{4M}+E_{1S} +q - \frac{k_2^2}{4M}-\frac{p_\ma{rel}^2}{M} \,,
\ee
where ${\bs k}_{1,2}$ are the c.m. momenta, ${\bs p}_{\ma{rel}}$ is the relative momentum between the unbound quark pair and $q=|{\bs q}|$ is the gluon energy. In the matrix element, $|\psi_{1S}\rangle$ is the hydrogen-like $1S$ wave function for the bound diquark in the anti-triplet, and $| \Psi_{{\bs p}_\ma{rel}} \rangle$ is the Coulomb wave function for the unbound sextet. The $1S$ binding energy is given by $E_{1S} = - \alpha_s^2M/9$. According to the power counting explained above, the c.m. kinetic energies will be neglected. Throughout this paper we set $\alpha_s = g^2/(4\pi)= 0.4$. 

To calculate rates, we need to average and sum over certain quantum numbers. For convenience, we define
\begin{widetext}
\be
|\ml{M}|^2 & \equiv & \sum_{a=1}^8 \sum_{l=1}^3 \sum_{\nu=1}^6 \sum_{\lambda = \pm}|\ml{M}^{\nu l a}_\lambda|^2 
= 2g^2q^2 |   \langle  \Psi_{{\bs p}_\ma{rel}}  | {\bs r} |\psi_\ma{1S}\rangle |^2\, 
\ee
\be
\label{eqn:reco}
\ml{F}^+ &\equiv& \frac{1}{2} g_+ \int\frac{\Diff{3}p_1}{(2\pi)^3}\frac{\Diff{3}p_2}{(2\pi)^3} \frac{\Diff{3}k_1}{(2\pi)^3} \frac{\Diff{3}q}{(2\pi)^32q} \big(1+n_B^{(q)}\big)  f_c({\bs x}, {\bs p}_1, t)f_c({\bs x}, {\bs p}_2, t) (2\pi)^4 \delta^3({\bs k}_1 + {\bs q} - {\bs k}_2) \delta(\Delta E) |\ml{M}|^2\\
\ml{F}^- &\equiv& \frac{1}{2} g_- \int \frac{\Diff{3}k_1}{(2\pi)^3}  \frac{\Diff{3}k_2}{(2\pi)^3}  \frac{\Diff{3}p_{\ma{rel}}}{(2\pi)^3} \frac{\Diff{3}q}{(2\pi)^32q}n_B^{(q)}  f_{cc}({\bs x}, {\bs k}_1, t)  (2\pi)^4 \delta^3({\bs k}_1 + {\bs q} - {\bs k}_2) \delta(\Delta E) |\ml{M}|^2 \, ,
\ee

\end{widetext}
where ${\bs p}_{\ma{rel}}$ and ${\bs k}_2$ are the relative and c.m. momenta of the unbound charm quark pair with momenta ${\bs p}_1$ and ${\bs p}_2$. The pre-factor $\frac{1}{2}$ avoids double counting in the phase space of two charm quarks. The $g$-factors are given by 
\be
g_+ &=& \frac{2J+1}{(2s+1)^2}\frac{d_6}{N_c^2}\frac{1}{d_6} = \frac{1}{12} \\
g_- &=& \frac{1}{d_{\bar{3}}} = \frac{1}{3}\,,
\ee
where $J=1$ is the diquark spin, $s=\frac{1}{2}$ is the heavy quark spin, $N_c = 3$ is the number of colors, $d_6 = 6$ is the sextet multiplicity and $d_{\bar{3}}=3$ is the anti-triplet multiplicity. For the formation process, one needs to average over the initial sextet multiplicity and only a fraction ${d_6}/{N_c^2}$ of unbound charm quark pairs are in the sextet, which can form a diquark by radiating out a gluon at the order of $r$ and $(1/M)^0$. The formed 1S diquark is a color anti-triplet and thus has to be in the spin triplet because of the antisymmetric nature of fermions. So another spin factor $\frac{2J+1}{(2s+1)^2} =\frac{3}{4}$ is inserted. For the dissociation process, one needs to average over the initial anti-triplet multiplicity.
The phase space measure is relativistic for gluons and non-relativistic for charm quarks and diquarks, which is consistent with our field definitions. 
Formation from unbound anti-triplet pairs only happens at higher orders in $r$ and $1/M$.

The gain and loss collision terms in the Boltzmann transport equations can be written as
\be
\label{eq:formterm}
\ml{C}_{c}^{\pm} &=& \frac{\delta \ml{F}^{\pm} }{\delta{{\bs p}_1}}  \bigg|_{{\bs p}_1 = {\bs p}} + \frac{\delta \ml{F}^{\pm} }{\delta{{\bs p}_2}}  \bigg|_{{\bs p}_2 = {\bs p}} \\
\ml{C}_{cc}^{\pm} &=& \frac{\delta \ml{F}^{\pm} }{\delta{{\bs k}_1}}  \bigg|_{{\bs k}_1 = {\bs p}} \,,
\ee
where the ``$\delta-$derivative" symbol is defined as
\be
&&\frac{\delta }{\delta{{\bs p}_i}} \int \prod_{j=1}^n \frac{\Diff{3}p_j}{(2\pi)^3} h({\bs p}_1, {\bs p}_2, \cdots, {\bs p}_n)\bigg|_{{\bs p}_i = {\bs p}} \\ \nn
&\equiv&  \frac{\delta }{\delta{a({\bs p}})} \int \prod_{j=1}^n \frac{\Diff{3}p_j}{(2\pi)^3} h({\bs p}_1, {\bs p}_2, \cdots, {\bs p}_n)a({\bs p}_i) \\ \nn
&=& \int \prod_{j=1, j\neq i}^n \frac{\Diff{3}p_j}{(2\pi)^3} h({\bs p}_1, {\bs p}_2, \cdots, {\bs p}_{i-1}, {\bs p}, {\bs p}_{i+1}, \cdots, {\bs p}_n)\,,
\ee
where the $\delta$ in the second line denotes the standard functional variation and $h({\bs p}_1, {\bs p}_2, \cdots, {\bs p}_n)$ and $a({\bs p}_i)$ are arbitrary independent functions.
In $\ml{C}_{c}^{\pm}$ two such ``$\delta-$derivatives" are involved because the initial or final states contain two charm quarks.

The rate of charm quarks combining $\Gamma_f$ and the dissociation rate of a diquark $\Gamma_d$ can be defined as
\be
\label{eq:formrate}
\ml{C}_{c}^{+} &\equiv& \Gamma_f({\bs x}, {\bs p}, t) f_{c}({\bs x}, {\bs p}, t) \\
\ml{C}_{cc}^{-} &\equiv& \Gamma_d({\bs x}, {\bs p}, t) f_{cc}({\bs x}, {\bs p}, t) \, .
\ee

The scattering amplitude and the rate are calculated in the rest frame of the diquark for dissociation and that of the unbound quark pair for formation, where the pNRQCD is valid. The Bose distribution of medium gluons $n_B^{(q)}$ is boosted into the rest frames, respectively. The two frames are not equivalent but since the gluon energy is small compared to $M$ ($T\ll M$), the difference is suppressed by $T/M$. We test the implementation of the formation and dissociation rates in a static QGP box with a constant temperature. After evolving for a sufficiently long period, the system of charm quarks and diquarks reaches thermal equilibrium. The equilibrium test is similar to that for heavy quarks and quarkonia \cite{Yao:2017fuc}.

To solve the transport equations, an initial condition is needed. Due to the large mass, the charm quark can be thought of being produced from the initial hard scattering in heavy ion collisions, before the QGP is formed. The initial transverse momentum and rapidity distribution from the hard scattering is calculated from FONLL \cite{FONLL} with the nuclear parton distribution function (PDF) EPS$09$ \cite{Eskola:2009uj}. The nuclear PDF contains a modification of the proton PDF due to nuclear many-body effects. The FONLL calculation is done with the renormalization and factorization scale $m_T=\sqrt{M^2+p_T^2}$. The number of charm quarks produced in one collision event is determined by $\sigma T_{AA}$, the product of the cross section $\sigma$ per binary collision calculated in FONLL, and the nuclear thickness function $T_{AA}$ derived from binary collision models. Here we will focus on collisions with $0-10\%$ centrality, which corresponds to impact parameters from $0$ to $5$ fm roughly and $T_{AA}\approx23$ mb \cite{Abelev:2013qoq}.

The initial position of the charm quark produced is sampled using the Trento model \cite{Moreland:2014oya}, a binary collision model. The model assumes the heavy ion collision is a superposition of a number of nucleon-nucleon collisions and calculates the spatial probability distribution where two nucleons from the approaching nuclei scatter. The charm quark production is a short-distance process, implying that its initial position is roughly the same as the location where the two parent nucleons scatter.

Each binary collision also deposits a certain amount of energy and entropy into the system. The Trento model also gives the initial energy and entropy densities. These are then fed into a $2+1$ dimensional viscous hydrodynamical simulator VISHNew \cite{Song:2007ux,Shen:2014vra}, which numerically solves the hydrodynamical equations 
\be
\partial_{\mu} T^{\mu\nu} &=& 0
\ee
with the energy-momentum tensor
\be
T^{\mu\nu} &=& eu^{\mu}u^{\nu} - (p+\Pi)(g^{\mu\nu}-u^{\mu}u^{\nu}) + \pi^{\mu\nu}, \\
\Pi &=& -\zeta\nabla\cdot u, \\
\pi^{\mu\nu} &=& 2\eta\nabla^{\langle\mu}u^{\nu\rangle}
\ee
for given initial conditions.
Here $e$ and $p$ are the local energy density and pressure, and $u^{\mu}$ is the local four-velocity of the QGP. $\Pi$ is the bulk stress with the bulk viscosity $\zeta$, and $\pi^{\mu\nu}$ is the shear stress tensor with the shear viscosity $\eta$. Here the angle bracket means traceless symmetrization.

With the initial condition and hydrodynamical background given, we solve the transport equations by test particles Monte Carlo simulations. The hydrodynamical simulation is assumed to start at the co-moving time $\tau=0.6$ fm/c. Before this, we assume the charm quarks are just free-streaming without interactions. After $\tau=0.6$ fm/c, we consider three types of processes at each time step $\Delta t=0.04$ fm/c in the laboratory frame: diffusion, formation and dissociation.

First, for each charm quark and diquark, we determine their thermal scattering rate with medium constituents. The product of the rate and time step $\Delta t$ gives the scattering probability. Then we use random numbers to determine whether a certain process occurs. If so, we sample the momenta of the incoming medium constituent from a thermal distribution and obtain the momenta of outgoing particles by energy-momentum conservation. Finally, we update both particles' momenta and positions after one time step.

Second, for each diquark, we calculate its dissociation rate and probability within a time step as above. If the diquark is determined to dissociate, we replace it by two unbound charm quarks whose momenta are determined from energy-momentum conservation and whose positions are given by that of the diquark just before the dissociation. 

Finally, for each charm quark with position ${\bs y}_i$ and momentum $\tilde{{\bs p}}_i$, whose neighboring charm quarks have positions ${\bs y}_j$ and momenta $\tilde{{\bs p}}_j$, we need to determine the diquark formation rate by using expressions (\ref{eqn:reco}), (\ref{eq:formterm}), (\ref{eq:formrate}). A problem appears, because the two quark distributions should be evaluated at the same position, but the product of two delta functions is ill-defined. We introduce a position dependence of the combination probability by means of a Gaussian function with a width chosen as the diquark Bohr radius $a_B=\alpha_sM/3$. This ensures that the combination rate for a widely separated charm quark pair vanishes. The product of the local distributions in (\ref{eqn:reco}) is thus replaced with
\begin{widetext}
\be
\label{eq:ff}
f_c({\bs x}, {\bs p}_1, t)f_{c}({\bs x}, {\bs p}_2, t) \rightarrow      
\sum_{i,j}  \frac{e^{-({\bs y}_i - {\bs y}_j)^2/2a_B^2}}{(2\pi a_B^2)^{3/2}} \delta^3\left({\bs x}-\frac{{\bs y}_i+{\bs y}_j}{2}\right)   
\delta^3({\bs p}_1-\tilde{{\bs p}}_i) \delta^3({\bs p}_2-\tilde{{\bs p}}_j)\,,
\ee
\end{widetext}
where the sum runs over all unbound charm quark pairs. For each charm quark $i$, the diquark formation rate in expression (\ref{eq:formrate}) involves a sum over $j$. If a diquark is formed, we replace the unbound charm quark pair by a diquark whose momentum is determined by momentum conservation and whose position is given by the center-of-mass position of the quark pair as indicated in (\ref{eq:ff}).

When particles reach the hadronization hypersurface determined by the local transition temperature $T_c\approx 154$ MeV, each diquark combines with a thermal up or down quark to form a doubly charmed baryon. Here we use a simple hadronization model: a massless up or down quark is sampled from a Fermi-Dirac distribution with the temperature $T_c$, and its momentum is added to the diquark momentum to determine the baryon momentum. The baryon energy is fixed by the momentum and vacuum mass $m(\Xi_{cc})$.  We assume all diquarks end up as the ground $\Xi_{cc}$ states because excited states decay to the ground state much faster than the weak decay of the ground state \cite{Li:2017pxa,Lu:2017meb}. In this way, roughly half the diquarks end up as $\Xi_{cc}^{++}$. A more realistic hadronization model would include the effect of the baryon wave function on the coalescence probability.

\begin{figure}
\centering
\vspace{0.1in}
\includegraphics[width=3.0in]{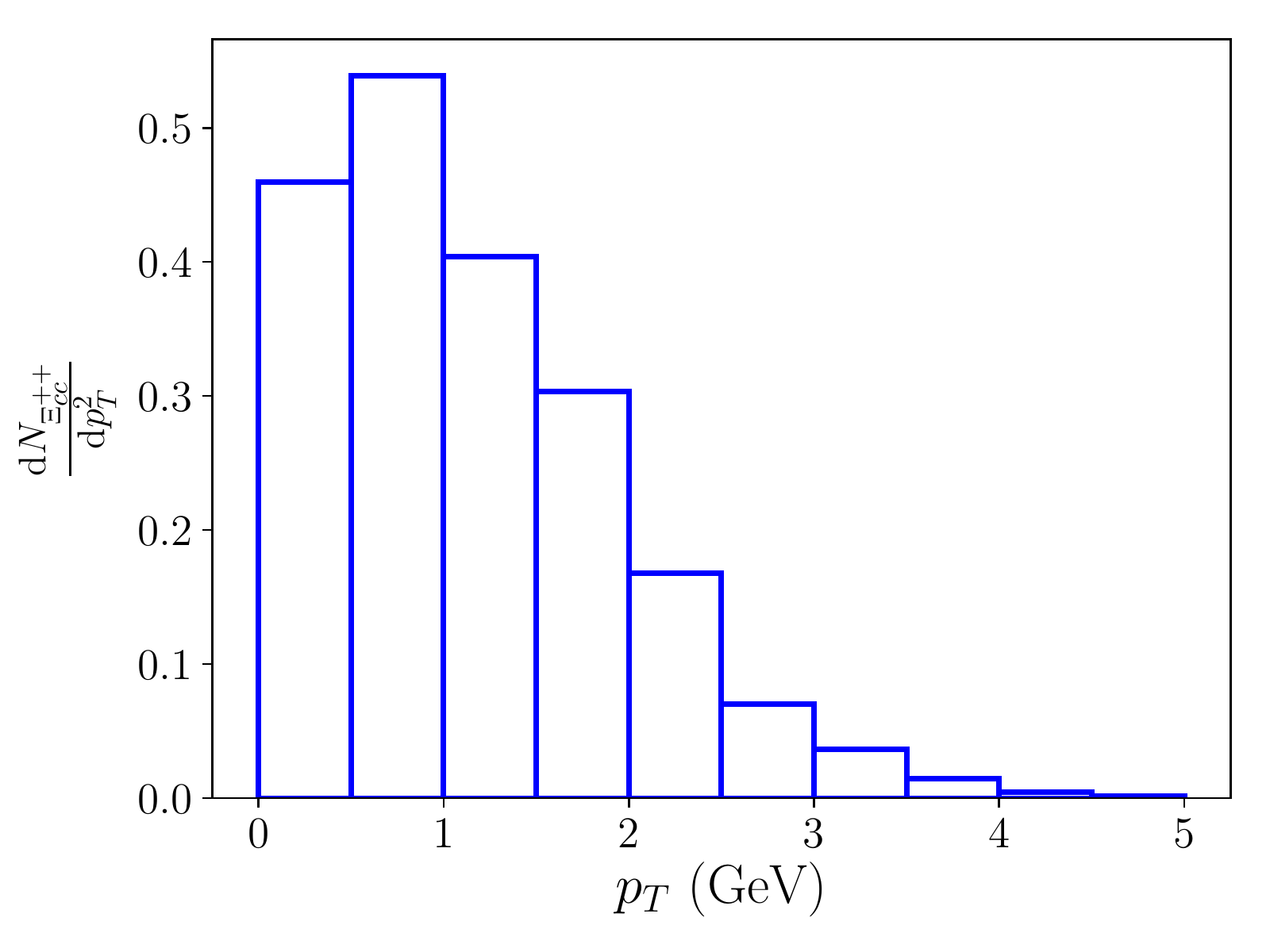}
\caption{$p_T$ spectra of emitted $\Xi_{cc}^{++}$ integrated over the rapidity window $-1 \leq y \leq 1$. The normalization is arbitrary.}\label{fig:pt}
\end{figure}

We have simulated 40,000 nuclear collision events. In each event, the initial charm quark momentum is sampled over the range $p_T \in [0,\,30]$ GeV and $y \in [-8,\,8]$. At the end of each calculation, we accept $\Xi_{cc}^{++}$ in the kinematic range $p_T \in [0,\,5]$ GeV and $y \in [-1,\,1]$. The $p_T$ spectra integrated over this rapidity range are shown in Fig.~\ref{fig:pt}. The yield within this kinematic range is $N(\Xi_{cc}^{++})\approx0.02$ per collision.

So far, we have assumed that the diquark can be formed at any temperature. This cannot be true due to the Debye screening of the attractive color force inside the QGP. To understand the influence of Debye screening on $\Xi_{cc}^{++}$ production, we repeat the calculation but assume a melting temperature $T_{\ma{m}}=250$ MeV above which the charm diquark cannot be formed inside the QGP. The yield in the same kinematic range is then reduced to $N(\Xi_{cc}^{++})\approx0.0125$ per collision.

The melting temperature of heavy diquarks can be studied from their free energies, in a similar way as quarkonia melting temperatures \cite{Mocsy:2013syh}. The free energy of a heavy quark pair could be studied on a lattice by calculating the correlations of two Polyakov loops at different lattice locations, where each Polyakov loop corresponds to a static thermal heavy quark \cite{McLerran:1981pb}. The free energy projected onto the color anti-triplet state can be used to study the binding energies and spectral functions of diquarks, from which one can obtain the melting temperature. The projections onto the anti-triplet and sextet states were first studied in Ref.~\cite{Nadkarni:1986as}. In the appendix, we explain how to project onto the anti-triplet in a gauge invariant but path dependent way. We also show that under a weak coupling expansion, the free energy of a pair of heavy quarks in the anti-triplet is the sum of the free energies of two individual heavy quarks and their attractive potential energy. A previous gauge dependent lattice study can be found in Ref.~\cite{Doring:2007uh}.

The calculation presented here can be improved in several ways. First, one can include higher-order corrections to the in-medium processes. The in-medium potentials of the diquark can also be made temperature-dependent by performing matching calculations between lattice results of Wilson loops and pNRQCD. Furthermore, one can use more realistic hadronization models. Finally, effects of the initial charm quark momentum distribution modifications from the pre-equilibrium effects could be studied.

The calculation can be extended to the production of other doubly heavy baryons, such as $\Xi_{bb}$ and $\Xi_{bc}$, and doubly heavy tetraquarks, among which the $bb\bar{u}\bar{d}$ ground state with $J^P=1^+$ is predicted to be stable \cite{Karliner:2017qjm,Eichten:2017ffp,Francis:2016hui,Bicudo:2016ooe,Czarnecki:2017vco}. The stability of heavy tetraquarks has been investigated previously in Ref.~\cite{Ader:1981db}. For $\Xi_{bb}$, the only difference is that fewer bottom quarks are produced than charm quarks. This implies that the probability of having two bottom quarks come close and form a bottom diquark is much smaller. Thus, one expects a correspondingly smaller yield of $\Xi_{bb}$. For $\Xi_{bc}$, there exist extra dipole terms in the pNRQCD Lagrangian for transitions among anti-triplets (or sextets) \cite{Brambilla:2005yk}, which means that an unbound pair of bottom and charm quarks in the anti-triplet channel can form a bound $bc$ diquark via a dipole transition.

For tetraquarks, the in-medium evolution of heavy quarks and diquarks proceeds in the same way, but the anti-triplet diquark hadronizes by coalescing with two light antiquarks. This process is analogous to the formation of an antibaryon containing a single heavy antiquark, while the formation of a doubly heavy baryon is analogous to the creation of a heavy meson. Heavy baryon ($\Lambda_c$) emission is known to be enhanced relative to heavy meson ($D^0$) emission in relativistic heavy ion collisions \cite{Zhou:2017ikn} as a consequence of quark recombination from the thermal quark-gluon plasma \cite{Oh:2009zj}, compared with proton-proton collisions. A similar enhancement of the production of doubly heavy tetraquarks, relative to the production of doubly heavy (anti-)baryons, can be expected. The measured ratio $\Lambda_c/D^0 \approx 1$ in Au+Au collisions at RHIC suggests that the yield of doubly heavy baryons and tetraquarks should also be approximately equal.

Finally, we discuss the feasibility of experimental measurements. The crucial factor is the yield-to-background ratio. Based on our calculations, the number of $\Xi_{cc}^{++}$ produced at the LHC energies may be large enough. But at the same time, higher collision energies mean higher levels of background. Though a measurement is currently difficult, it is promising that the noisy background difficulty will be overcome in the future with detector upgrades such as the ALICE Inner Tracking System upgrade. With the high-resolution detectors, one can apply stricter topological cuts to reduce the level of background and increase the yield-to-background ratio. Just as the STAR Collaboration first measured the $\Lambda_c$ production in heavy ion collisions with the newly installed Heavy Flavor Tracker \cite{Zhou:2017ikn}, measurements of doubly heavy baryons and even bound tetraquarks in heavy ion collisions may become possible in the future. Experimental measurements rely on the reconstruction from decay products of $\Xi_{cc}^{++}$. The decay properties of doubly heavy baryons have been intensely studied \cite{SanchisLozano:1993kh,Kiselev:1998sy,Guberina:1999mx,Egolf:2002nk,Ebert:2004ck,Hu:2005gf,Albertus:2011xz,Karliner:2014gca,Wang:2017mqp}.

In conclusion, we have used Boltzmann transport equations to describe the in-medium formation, dissociation, and diffusion of charm diquarks. Based on it, we estimate the production rate and $p_T$ spectra of the doubly charmed baryon $\Xi_{cc}^{++}$ in central Pb-Pb collision at $2.76$ TeV. It will be of great interest if experimental efforts are taken to try to measure $\Xi_{cc}^{++}$ in heavy ion collisions. A measurement of the production rate would allow us to extract the melting temperature of the charm diquark in QGP from the above calculation. Comparison can be made with the melting temperature calculated from lattice results of the free energy of the anti-triplet. These experimental and lattice studies would provide valuable information to our understanding of QCD at finite temperature and properties of QGP.

\begin{acknowledgments}
We acknowledge stimulating communications with Marek Karliner. X.Y. thanks Weiyao Ke and Chun Shen for helpful discussions and the nuclear theory group at Brookhaven National Laboratory, where part of this work was completed, for its hospitality. The work is supported from U.S. Department of Energy under Research Grant No. DE-FG02-05ER41367. X.Y. also acknowledges support from Brookhaven National Laboratory.
\end{acknowledgments}

\begin{widetext}
\appendix*
\section{Free energy of a heavy quark pair in the anti-triplet and sextet}
The anti-triplet and sextet states of a heavy quark pair at different lattice locations can be defined as
\be
| QQ_{\bar{3}}({\bs 0}, {\bs r}, \tau)\rangle^{l} &\equiv& \frac{1 }{\sqrt{2}}\epsilon_{ikl}\psi_i^{\dagger}({\bs 0},\tau)  \psi_j^{\dagger}({\bs r},\tau) W^{\dagger}_{jk} ( ({\bs 0},\tau), ({\bs r},\tau))|s\rangle \\
| QQ_{6}({\bs 0}, {\bs r}, \tau)\rangle^{\nu} &\equiv& \sigma^{\nu}_{ik} \psi_i^{\dagger}({\bs 0},\tau)  \psi_j^{\dagger}({\bs r},\tau) W^{\dagger}_{jk} ( ({\bs 0},\tau), ({\bs r},\tau))|s\rangle\,,
\ee
where $\tau$ is the Euclidean time and $|s\rangle$ can be any state with no heavy quarks. The symbol $\sigma^{\nu}_{ik}$ is defined in the expressions (\ref{eqn:six1}) and (\ref{eqn:six2}) and satisfies $\sigma^{\nu}_{ik}\sigma^{\nu}_{i'k'} = (\delta_{ii'}\delta_{kk'} + \delta_{ik'}\delta_{i'k})/2$. The symbol $W(y,x)$ denotes a Wilson line from lattice site $x$ to site $y$. The definitions depend on the spatial path of the Wilson line. The heavy quark annihilation $\psi$ and creation $\psi^{\dagger}$ operators satisfy the anti-commutation relation on the lattice
\be
\{  \psi_{i}({\bs r},\tau) , \psi_j^{\dagger}({\bs r'},\tau)   \} = \delta_{{\bs r}{\bs r'}} \delta_{ij}\,.
\ee

The free energy of a heavy quark pair in the anti-triplet can be defined as
\be
e^{-F_{QQ(\bar{3})}(\bs r)/T} &=& \frac{1}{N_c}\sum_{|s\rangle}    \langle QQ_{\bar{3}}({\bs 0}, {\bs r}, 0)|^{l}  e^{-\beta H}  | QQ_{\bar{3}}({\bs 0}, {\bs r}, 0)\rangle^{l} \\ \nn
&=&\frac{1}{2N_c}\epsilon_{i'k'l}\epsilon_{ikl}\sum_{|s\rangle} \langle s | W_{k'j'}( ({\bs 0},0), ({\bs r},0) )\psi_{j'}({\bs r},0) \psi_{i'}({\bs 0},0) e^{- \beta H} \psi_i^{\dagger}({\bs 0},0)  \psi_j^{\dagger}({\bs r},0) W^{\dagger}_{jk} ( ({\bs 0},0), ({\bs r},0) )|s\rangle \\\nn
&=&\frac{1}{6} (\delta_{ii'}\delta_{kk'} - \delta_{ik'}\delta_{i'k}) \sum_{|s\rangle} \langle s |  e^{- \beta H}  W_{k'j'}( ({\bs 0},\beta), ({\bs r},\beta) )\psi_{j'}({\bs r},\beta) \psi_{i'}({\bs 0},\beta) \psi_i^{\dagger}({\bs 0},0)  \psi_j^{\dagger}({\bs r},0) W^{\dagger}_{jk} ( ({\bs 0},0), ({\bs r},0) )   |s\rangle \,.
\ee 
In the static heavy quark limit \cite{McLerran:1981pb},
\be
\psi_i ({\bs r},\beta) = \ml{T}(e^{ig\int_0^{\beta} \diff\tau A_0(\bs r, \tau)})_{ij} \psi_j ({\bs r},0) \equiv L(\bs r)_{ij} \psi_j ({\bs r},0)\,,
\ee 
where $\ml{T}$ is the time ordering operator. The starting and ending points of the Wilson line along the Euclidean time direction are the same due to the periodicity of gauge fields at finite temperature and is denoted as the Polyakov line $L(\bs r)$. Then using the anti-commutation relation of heavy quark operators it can be shown
\be
e^{-F_{QQ(\bar{3})}(\bs r)/T} = \frac{1}{6} \langle  \Tr L({\bs 0}) \Tr L({\bs r})  \rangle_T -\frac{1}{6} \langle   \Tr[W( ({\bs 0},\beta), ({\bs r},\beta)) L({\bs r})  W^{\dagger}( ({\bs 0},0), ({\bs r},0) )   L({\bs 0})      ]    \rangle_T\,,
\ee
where $\langle \hat{O} \rangle_T \equiv \sum_{|s\rangle} \langle s |e^{-\beta H} \hat{O} |s\rangle $ and $\Tr L$ is the Polyakov loop. Both the correlation terms in the above expression are gauge invariant because of the cyclic property of the trace and the periodicity of gauge fields. Schematic diagrams for the two correlation terms are shown in Fig.~\ref{fig:lattice}. 

\begin{figure}
\centering
\subfigure[]{\includegraphics[width=.4\textwidth]{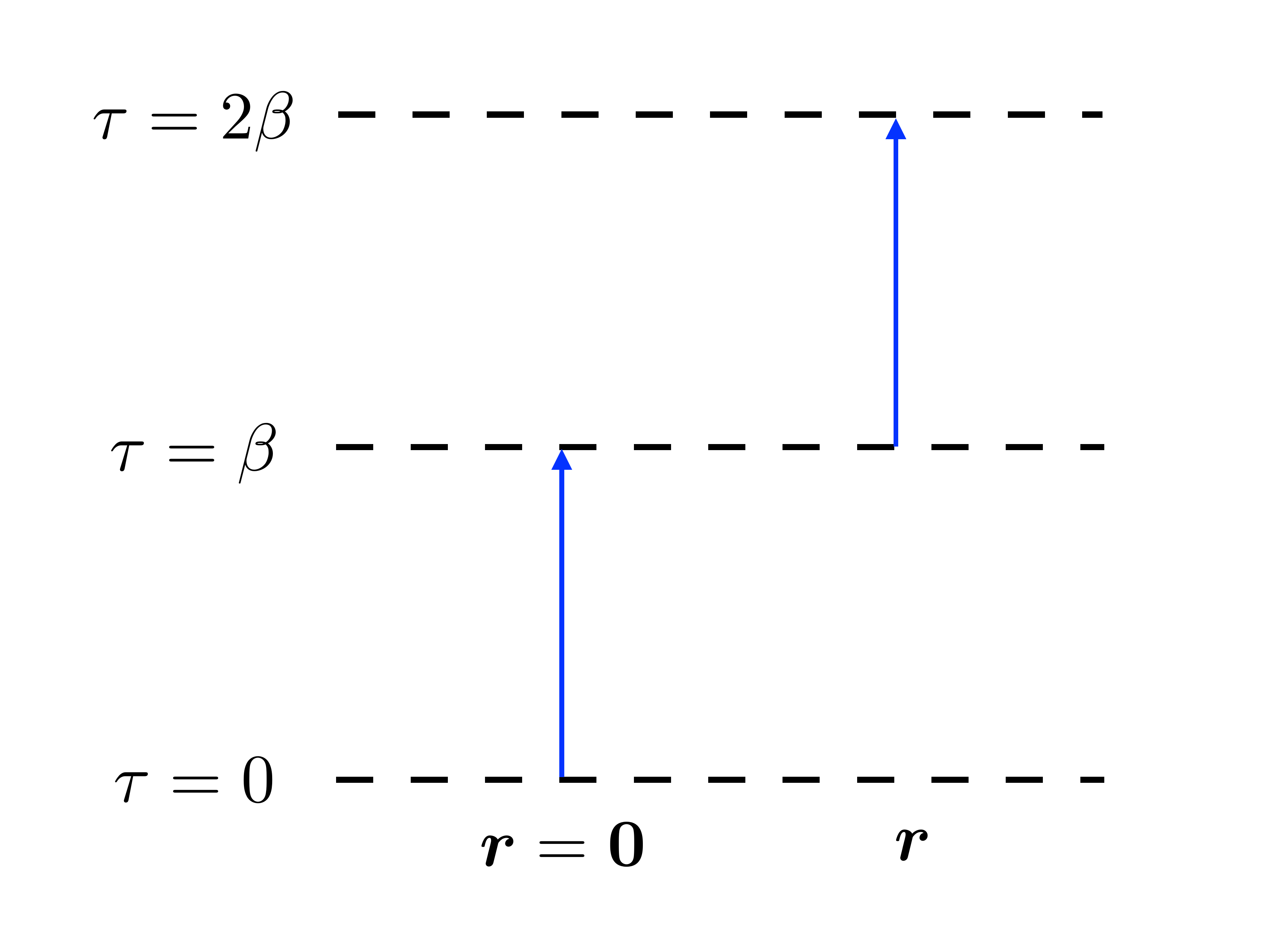}}
\subfigure[]{\includegraphics[width=.4\textwidth]{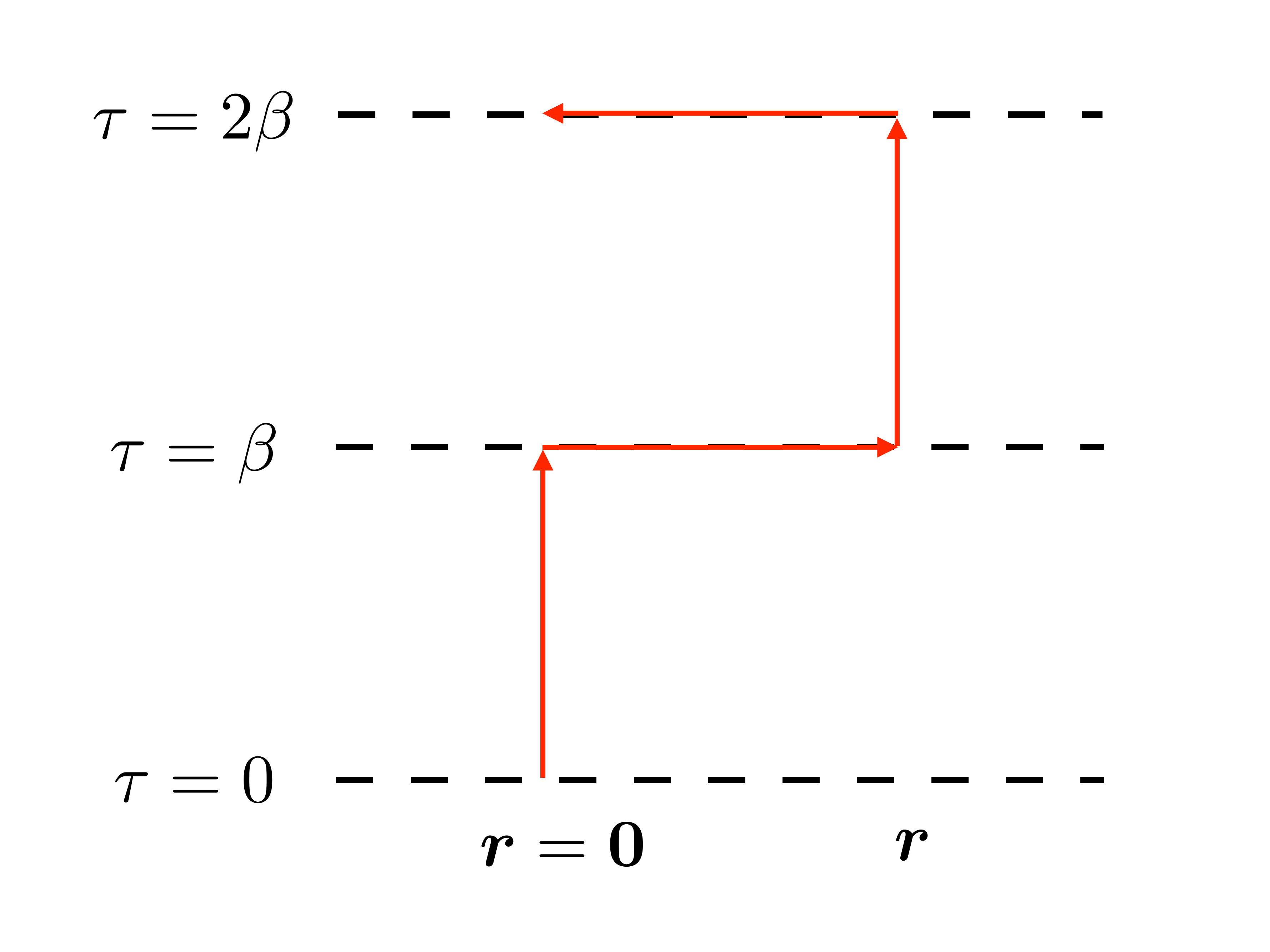}}
\caption{Schematic diagrams for the correlation terms in $\exp(-F_{QQ(\bar{3})}(\bs r)/T)$. The sub-plots (a) and (b) correspond to the first and second terms separately. The three black dashed lines label the same Euclidean time due to the periodicity at finite temperature. The region from $\tau=\beta$ to $\tau =2\beta$ is just a duplicate of the region from $\tau=0$ to $\tau = \beta$. In (a), the two blue arrows indicate the two Polyakov loops which are located at $\bs r = 0$ and $\bs r$. In (b), the four red arrows indicate the trace in the second term. It consists of a Polyakov line at $\bs r =0$, followed by a Wilson line from $\bs r = 0$ to $\bs r$, then another Polyakov line at $\bs r$ and finally a Wilson line from $\bs r$ to $\bs r = 0$. Though straight lines are used to denote the Wilson lines, they can be any spatial paths connecting the two ends. Due to the periodicity of gauge fields, the four red arrows form a loop.}
\label{fig:lattice}
\end{figure}

In a similar way, the sextet free energy can be defined as
\be
e^{-F_{QQ(6)}(\bs r)/T} &=& \frac{1}{6}\sum_{|s\rangle}    \langle QQ_{6}({\bs 0}, {\bs r}, 0)|^{\nu}  e^{-\beta H}  | QQ_{6}({\bs 0}, {\bs r}, 0)\rangle^{\nu} \\ 
&=&\frac{1}{12} \langle  \Tr L({\bs 0}) \Tr L({\bs r})  \rangle_T + \frac{1}{12} \langle   \Tr[W( ({\bs 0},\beta), ({\bs r},\beta)) L({\bs r})  W^{\dagger}( ({\bs 0},0), ({\bs r},0) )   L({\bs 0})      ]    \rangle_T\,,
\ee
which is also gauge invariant. Both definitions depend on the spatial paths of the Wilson lines.

Under a weak coupling expansion in powers of $g$, we obtain in the static gauge $\dot{A}_0=0$ (where $A_0 $ is a constant matrix)
\be
e^{-F_{QQ(\bar{3})}({\bs r})/T} = 1 + \frac{g^2\beta^2}{12}\delta^{ab} \langle  A_0^{a}(\bs r)A_0^{b}(\bs 0)\rangle_T - \frac{g^2\beta^2}{12}\delta^{ab} \langle  A_0^{a}(\bs 0)A_0^{b}(\bs 0)\rangle_T - \frac{g^2\beta^2}{12}\delta^{ab} \langle  A_0^{a}(\bs r)A_0^{b}(\bs r)\rangle_T +\ml{O}(g^3)\,.
\ee
The last two terms are independent of the positions and are just the free energies of two individual heavy quarks at order $g^2$. The free energy of a single heavy quark can be calculated from
\be
e^{-F_Q/T} = \frac{1}{3} \langle \Tr L  \rangle_T \,.
\ee
Therefore,
\be
F_{QQ(\bar{3})}({\bs r})  = 2F_Q - \frac{g^2\beta}{12}\delta^{ab} \langle  A_0^{a}(\bs r)A_0^{b}(\bs 0)\rangle_T+ \ml{O}(g^3)\,.
\ee
In the static gauge and under the hard thermal loop approximation
\be
\langle  A_0^{a}(\bs r)A_0^{b}(\bs 0)\rangle_T = T\sum_n\int \frac{\diff^3q}{(2\pi)^3}\frac{e^{i{\bs q}\cdot{\bs r} } }{{\bs q}^2+m_D^2}\delta_{n0} \delta^{ab}
= T\delta^{ab} \frac{1}{4\pi r}e^{-m_Dr}\,.
\ee
So finally,
\be
F_{QQ(\bar{3})}({\bs r})  = 2F_Q - \frac{2}{3} \frac{g^2}{4\pi r}e^{-m_Dr} + \ml{O}(g^3)\,.
\ee
The free energy of an anti-triplet heavy quark pair is the sum of the free energies of two individual heavy quarks and their color attractive potential energy.

In a similar way,
\be
F_{QQ(6)}({\bs r})  = 2F_Q + \frac{1}{3} \frac{g^2}{4\pi r}e^{-m_Dr} + \ml{O}(g^3)\,.
\ee
The free energy of a sextet is the sum of the free energies of two individual heavy quarks and their color repulsive potential energy. Though up to order $g^2$ the anti-triplet and sextet free energies are independent of the Wilson line paths in the definition, they are generally dependent on the paths beyond the leading order \cite{Nadkarni:1986cz}.\\

\end{widetext}

\bibliographystyle{apsrev4-1}

\begin{thebibliography}{99}
\bibitem{Aaij:2017ueg} 
  R.~Aaij {\it et al.} [LHCb Collaboration],
  Phys.\ Rev.\ Lett.\  {\bf 119}, no. 11, 112001 (2017)
  [arXiv:1707.01621 [hep-ex]].

\bibitem{Mattson:2002vu} 
  M.~Mattson {\it et al.} [SELEX Collaboration],
  Phys.\ Rev.\ Lett.\  {\bf 89}, 112001 (2002)
  [hep-ex/0208014];
  A.~Ocherashvili {\it et al.} [SELEX Collaboration],
  Phys.\ Lett.\ B {\bf 628}, 18 (2005)
  [hep-ex/0406033].
     
\bibitem{Karliner:2014gca} 
  M.~Karliner and J.~L.~Rosner,
  Phys.\ Rev.\ D {\bf 90}, no. 9, 094007 (2014)
  [arXiv:1408.5877 [hep-ph]].

\bibitem{Becattini:2005hb} 
  F.~Becattini,
  Phys.\ Rev.\ Lett.\  {\bf 95}, 022301 (2005)
  [hep-ph/0503239].
  
\bibitem{Zhao:2016ccp} 
  J.~Zhao, H.~He and P.~Zhuang,
  Phys.\ Lett.\ B {\bf 771}, 349 (2017)
  [arXiv:1603.04524 [nucl-th]].

\bibitem{Adamczyk:2017xur} 
  L.~Adamczyk {\it et al.} [STAR Collaboration],
  Phys.\ Rev.\ Lett.\  {\bf 118}, no. 21, 212301 (2017)
  [arXiv:1701.06060 [nucl-ex]].
    
\bibitem{Yao:2017fuc} 
  X.~Yao and B.~M\"uller,
  Phys.\ Rev.\ C {\bf 97}, no. 1, 014908 (2018)
  [arXiv:1709.03529 [hep-ph]].
  
\bibitem{Gossiaux:2008jv} 
  P.~B.~Gossiaux and J.~Aichelin,
  Phys.\ Rev.\ C {\bf 78}, 014904 (2008)
  [arXiv:0802.2525 [hep-ph]].

\bibitem{Gossiaux:2009mk} 
  P.~B.~Gossiaux, R.~Bierkandt and J.~Aichelin,
  Phys.\ Rev.\ C {\bf 79}, 044906 (2009)
  [arXiv:0901.0946 [hep-ph]].
 
\bibitem{Uphoff:2014hza} 
  J.~Uphoff, O.~Fochler, Z.~Xu and C.~Greiner,
  J.\ Phys.\ G {\bf 42}, no. 11, 115106 (2015)
  [arXiv:1408.2964 [hep-ph]].
  
\bibitem{duke_lbt} 
  S.~Bass, W.~Ke and Y.~Xu,
  (to be published).
   
\bibitem{Brambilla:2005yk} 
  N.~Brambilla, A.~Vairo and T.~Rosch,
  Phys.\ Rev.\ D {\bf 72}, 034021 (2005)
  [hep-ph/0506065].

\bibitem{Fleming:2005pd} 
  S.~Fleming and T.~Mehen,
  Phys.\ Rev.\ D {\bf 73}, 034502 (2006)
  [hep-ph/0509313].

\bibitem{Brambilla:2008cx} 
  N.~Brambilla, J.~Ghiglieri, A.~Vairo and P.~Petreczky,
  Phys.\ Rev.\ D {\bf 78}, 014017 (2008)
  [arXiv:0804.0993 [hep-ph]];
  N.~Brambilla, M.~A.~Escobedo, J.~Ghiglieri and A.~Vairo,
  JHEP {\bf 1112}, 116 (2011)
  [arXiv:1109.5826 [hep-ph]];
  N.~Brambilla, M.~A.~Escobedo, J.~Ghiglieri and A.~Vairo,
  JHEP {\bf 1305}, 130 (2013)
  [arXiv:1303.6097 [hep-ph]].
  
\bibitem{FONLL}
  M.~Cacciari, M.~Greco and P.~Nason,
  JHEP {\bf 9805} (1998) 007 [arXiv:hep-ph/9803400];
  M.~Cacciari, S.~Frixione and P.~Nason,
  JHEP {\bf 0103} (2001) 006 [arXiv:hep-ph/0102134].

\bibitem{Eskola:2009uj} 
  K.~J.~Eskola, H.~Paukkunen and C.~A.~Salgado,
  JHEP {\bf 0904}, 065 (2009)
  [arXiv:0902.4154 [hep-ph]].


\bibitem{Abelev:2013qoq} 
  B.~Abelev {\it et al.} [ALICE Collaboration],
  Phys.\ Rev.\ C {\bf 88}, no. 4, 044909 (2013)
  [arXiv:1301.4361 [nucl-ex]].
  
\bibitem{Moreland:2014oya} 
  J.~S.~Moreland, J.~E.~Bernhard and S.~A.~Bass,
  Phys.\ Rev.\ C {\bf 92}, no. 1, 011901 (2015)
  [arXiv:1412.4708 [nucl-th]].

\bibitem{Song:2007ux} 
  H.~Song and U.~W.~Heinz,
  Phys.\ Rev.\ C {\bf 77}, 064901 (2008)
  [arXiv:0712.3715 [nucl-th]].
        
\bibitem{Shen:2014vra} 
  C.~Shen, Z.~Qiu, H.~Song, J.~Bernhard, S.~Bass and U.~Heinz,
  Comput.\ Phys.\ Commun.\  {\bf 199}, 61 (2016)
  [arXiv:1409.8164 [nucl-th]].

\bibitem{Li:2017pxa} 
  H.~S.~Li, L.~Meng, Z.~W.~Liu and S.~L.~Zhu,
  Phys.\ Lett.\ B {\bf 777}, 169 (2018)
  [arXiv:1708.03620 [hep-ph]].
  
\bibitem{Lu:2017meb} 
  Q.~F.~L\"u, K.~L.~Wang, L.~Y.~Xiao and X.~H.~Zhong,
  Phys.\ Rev.\ D {\bf 96}, no. 11, 114006 (2017)
  [arXiv:1708.04468 [hep-ph]].

\bibitem{Mocsy:2013syh} 
  A.~Mocsy, P.~Petreczky and M.~Strickland,
  Int.\ J.\ Mod.\ Phys.\ A {\bf 28}, 1340012 (2013)
  [arXiv:1302.2180 [hep-ph]].

\bibitem{McLerran:1981pb} 
  L.~D.~McLerran and B.~Svetitsky,
  Phys.\ Rev.\ D {\bf 24}, 450 (1981).
 
\bibitem{Nadkarni:1986as}
  S.~Nadkarni,
  Phys.\ Rev.\ D {\bf 34} (1986) 3904.
  
\bibitem{Doring:2007uh} 
  M.~Doring, K.~Huebner, O.~Kaczmarek and F.~Karsch,
  Phys.\ Rev.\ D {\bf 75}, 054504 (2007)
  [hep-lat/0702009].
  
\bibitem{Karliner:2017qjm} 
  M.~Karliner and J.~L.~Rosner,
  Phys.\ Rev.\ Lett.\  {\bf 119}, no. 20, 202001 (2017)
  [arXiv:1707.07666 [hep-ph]].
  
\bibitem{Eichten:2017ffp} 
  E.~J.~Eichten and C.~Quigg,
  Phys.\ Rev.\ Lett.\  {\bf 119}, no. 20, 202002 (2017)
  [arXiv:1707.09575 [hep-ph]].

\bibitem{Francis:2016hui} 
  A.~Francis, R.~J.~Hudspith, R.~Lewis and K.~Maltman,
  Phys.\ Rev.\ Lett.\  {\bf 118}, no. 14, 142001 (2017)
  [arXiv:1607.05214 [hep-lat]].
 
\bibitem{Bicudo:2016ooe} 
  P.~Bicudo, J.~Scheunert and M.~Wagner,
  Phys.\ Rev.\ D {\bf 95}, no. 3, 034502 (2017)
  [arXiv:1612.02758 [hep-lat]].

\bibitem{Czarnecki:2017vco} 
  A.~Czarnecki, B.~Leng and M.~B.~Voloshin,
  arXiv:1708.04594 [hep-ph].

\bibitem{Ader:1981db} 
  J.~P.~Ader, J.~M.~Richard and P.~Taxil,
  Phys.\ Rev.\ D {\bf 25}, 2370 (1982).
        
\bibitem{Zhou:2017ikn} 
  L.~Zhou [STAR Collaboration],
  Nucl.\ Phys.\ A {\bf 967}, 620 (2017)
  [arXiv:1704.04364 [nucl-ex]].
  
\bibitem{Oh:2009zj} 
  Y.~Oh, C.~M.~Ko, S.~H.~Lee and S.~Yasui,
  Phys.\ Rev.\ C {\bf 79}, 044905 (2009)
  [arXiv:0901.1382 [nucl-th]].

\bibitem{SanchisLozano:1993kh} 
  M.~A.~Sanchis-Lozano,
  Phys.\ Lett.\ B {\bf 321}, 407 (1994);
  Nucl.\ Phys.\ B {\bf 440}, 251 (1995)
  [hep-ph/9502359].

\bibitem{Kiselev:1998sy} 
  V.~V.~Kiselev, A.~K.~Likhoded and A.~I.~Onishchenko,
  Phys.\ Rev.\ D {\bf 60}, 014007 (1999)
  [hep-ph/9807354].

\bibitem{Guberina:1999mx} 
  B.~Guberina, B.~Melic and H.~Stefancic,
  Eur.\ Phys.\ J.\ C {\bf 9}, 213 (1999)
  [Eur.\ Phys.\ J.\ C {\bf 13}, 551 (2000)]
  [hep-ph/9901323].

\bibitem{Egolf:2002nk} 
  D.~A.~Egolf, R.~P.~Springer and J.~Urban,
  Phys.\ Rev.\ D {\bf 68}, 013003 (2003)
  [hep-ph/0211360].

\bibitem{Ebert:2004ck} 
  D.~Ebert, R.~N.~Faustov, V.~O.~Galkin and A.~P.~Martynenko,
  Phys.\ Rev.\ D {\bf 70}, 014018 (2004)
  Erratum: [Phys.\ Rev.\ D {\bf 77}, 079903 (2008)]
  [hep-ph/0404280].

\bibitem{Hu:2005gf} 
  J.~Hu and T.~Mehen,
  Phys.\ Rev.\ D {\bf 73}, 054003 (2006)
  [hep-ph/0511321];
  T.~Mehen,
  Phys.\ Rev.\ D {\bf 96}, no. 9, 094028 (2017)
  [arXiv:1708.05020 [hep-ph]].
  
\bibitem{Albertus:2011xz} 
  C.~Albertus, E.~Hern\'andez and J.~Nieves,
  Phys.\ Lett.\ B {\bf 704}, 499 (2011)
  [arXiv:1108.1296 [hep-ph]].

\bibitem{Wang:2017mqp} 
  W.~Wang, F.~S.~Yu and Z.~X.~Zhao,
  Eur.\ Phys.\ J.\ C {\bf 77}, 781 (2017)
  [arXiv:1707.02834 [hep-ph]];
  W.~Wang, Z.~P.~Xing and J.~Xu,
  Eur.\ Phys.\ J.\ C {\bf 77}, 800 (2017)
  [arXiv:1707.06570 [hep-ph]].

\bibitem{Nadkarni:1986cz} 
  S.~Nadkarni,
  Phys.\ Rev.\ D {\bf 33}, 3738 (1986).
       
\end{thebibliography}

\end{document}